\documentclass[useAMS,usenatbib,usegraphicx]{mn2e}

\title[Collimated outflows in the planetary nebula Hu\,1-2]{The collimated
  outflows of the planetary nebula Hu\,1-2: proper motion and radial velocity
  measurements\thanks{Based on observations made with the Nordic Optical
    Telescope (NOT) which is operated jointly by Denmark, Finland, Iceland, Norway, and
  Sweden on the island of La Palma in the Spanish Observatorio del Roque de
  los Muchachos of the Instituto de Astrof\'{\i}sica de Canarias (IAC).}}
\author[L.F. Miranda et al.]{L.F. Miranda$^{1,2}$\thanks{E-mail:
lfm@iaa.es (LFM); blanco@iaa.es (MB); mar@iaa.es (MAR); angles.riera@upc.edu (AR)}, M. Blanco$^{3}$\footnotemark[2], 
M.A. Guerrero$^{3}$\footnotemark[2], A. Riera$^{4}$\footnotemark[2] \\
$^{1}$ Departamento de F\'{\i}sica Aplicada, Facultade de Ciencias, Campus Lagoas-Marcosende s/n, 
Universidade de Vigo, \\ E-36310 Vigo, Spain (present address) \\
$^{2}$ Consejo Superior de Investigaciones Cient\'{\i}ficas, C/ Serrano 117, E-28006 Madrid, Spain \\
$^{3}$ Instituto de Astrof\'{\i}sica de Andaluc\'{\i}a -- CSIC, C/ Glorieta de la Astronom\'{\i}a s/n, 
E-18008 Granada, Spain \\
$^{4}$ Departament de F\'{\i}sica i Enginyeria Nuclear, EUETIB, Universitat Polit\`ecnica de Catalunya, Compte 
d'Urgell 187, \\ E-08036 Barcelona, Spain 
}

\begin{document}

\date{Accepted. Received; in original form}

\pagerange{\pageref{firstpage}--\pageref{lastpage}} \pubyear{}

\maketitle

\label{firstpage}

\begin{abstract}
  Hu\,1-2 is a planetary nebula that contains an isolated knot located northwestern of the main nebula, which could be 
  related to a collimated outflow. We present a subsarcsecond H$\alpha$+[N\,{\sc ii}] image and a high-resolution,
  long-slit spectrum of Hu\,1-2 that allow us to identify the southeastern counterpart of the northwestern knot and 
  to establish their high velocity ($>$ 340 km\,s$^{-1}$), collimated bipolar outflow nature. The detection of the 
  northwestern knot
  in POSS red plates allows us to carry out a proper motion analysis by combining three POSS red plates and 
  two narrow-band H$\alpha$+[N\,{\sc ii}] CCD images, with a time baseline of $\simeq$ 57 yr. A proper motion 
  of 20 $\pm$ 6\,mas\,yr$^{-1}$ along position angle 312$^{\circ}$ $\pm$ 15$^{\circ}$, and a dynamical age
  of 1375$^{+590}$\llap{$_{-320}$}\,yr are obtained for the bipolar
  outflow. The measured proper motion and the spatio-kinematical 
  properties of the bipolar outflow yield a lower limit of 2.7\,kpc for the distance to Hu\,1-2.
\end{abstract}


\section{Introduction}

Planetary nebulae (PNe) represent the final evolutionary stages of low-- and
intermediate--mass stars. They are formed when the envelope ejected during the
previous Asymptotic Giant Branch phase is ionized by the hot central star. In the 
last years the presence of collimated bipolar outflows in PNe has been widely recognized. 
These outflows probably play a crucial role in the shaping and dynamical evolution of 
PNe \citep{b25}. Therefore, the knowledge of their dynamical properties is critical to understand how PNe form. The 
kinematics of the outflows has been studied by means of high-resolution spectroscopy that 
permits to obtain the radial component of the velocity (e.g., Miranda et al. 1999, 2001; Guerrero et al. 
2000). The measurement of the tangential component, i.e., 
the proper motion of collimated outflows, is restricted to a very few cases and it has been obtained 
mostly from {\it HST} observations (NGC\,7009: Fern\'andez et al. 2004; He\,2-90: Sahai et al. 2002;
Hen\,3-1475: Borkowski \& Harrington 2001; Riera et al. 2003) or {\it VLA} radio
continuum data (NGC\,7009: Rodr\'{\i}guez \& G\'omez 2007) with the noticeable
exceptions of the works by Liller (1965) and Meaburn (1997) who used photographic material to measure
proper motions of the collimated outflows in NGC\,7009 and KjPn\,8,
respectively. The measurement of proper motions is important because it allows us to
obtain the distances to the PN, if the expansion velocity of the outflows is
known (e.g., Meaburn 1997), or the expansion velocity vector, if the distance is known 
(e.g., Fern\'andez et al.2007). Therefore, further measurements of proper motions of
collimated outflows in PNe are highly desirable. 

Collimated outflows may exist in the PN Hu\,1-2 (PN\,G086.5$-$08.8) that has been classified as 
elliptical with ansae \citep{b4}. The kinematics of the inner nebular 
regions ($\simeq$ 10$''$) of Hu\,1-2 was analyzed by Sabbadin et al. (1983,
1987). These authors identified a toroid and bipolar lobes but a
reconstruction of the spatio-kinematical structure was not possible because of the
peculiar velocity field. In addition, one single knot can be hinted in the images by Manchado et al. (1996) 
northwestwards and well outside the main nebula, while a possible southeastern
counterpart is superimposed by a field star. 
Remarkably, the northwestern knot can be identified in POSS red plates of the Digitized Sky Surveys. In this
respect, Hu\,1-2 constitutes a rare case (together with NGC\,7009 and KjPn\,8) 
because knots/collimated outflows in PNe usually are weak and/or located close to the much
brighter main nebula, which do not favor their detection in POSS
plates. Moreover, the H$\alpha$ and [N\,{\sc ii}] emission lines are by far the dominant 
emissions from knots/collimated outflows in PNe in the optical (e.g., Balick 1993, 1994), 
as it is the case of the northwestern knot of Hu\,1-2. In consequence, the emission from the 
northwestern knot of Hu\,1-2 detected in the POSS red plates can be 
attributed to these two emission lines. Therefore, Hu\,1-2 offers an 
excellent opportunity to attempt a proper motion analysis of knots/collimated outflows in PNe 
by combining POSS red plates with modern H$\alpha$+[N\,{\sc ii}] imagery.

In this paper we present a new H$\alpha$+[N\,{\sc ii}] image of Hu\,1-2
obtained under subarsecond conditions, and a high resolution, long-slit spectrum 
that allow us to identify the southeastern counterpart of the northwestern knot and to establish that 
these two knots constitute a high velocity, collimated bipolar outflow. In addition, we carry out 
an analysis of five images obtained at different epochs, including three images from the POSS, 
to measure the proper motion of the northwestern knot and to constrain the distance to Hu\,1-2. 

\section{Observations}

\subsection{Direct images}

The five images of Hu\,1-2 analyzed in this paper are the following:

\begin{figure}
   \begin{center}
   \includegraphics[width=80mm,clip=]{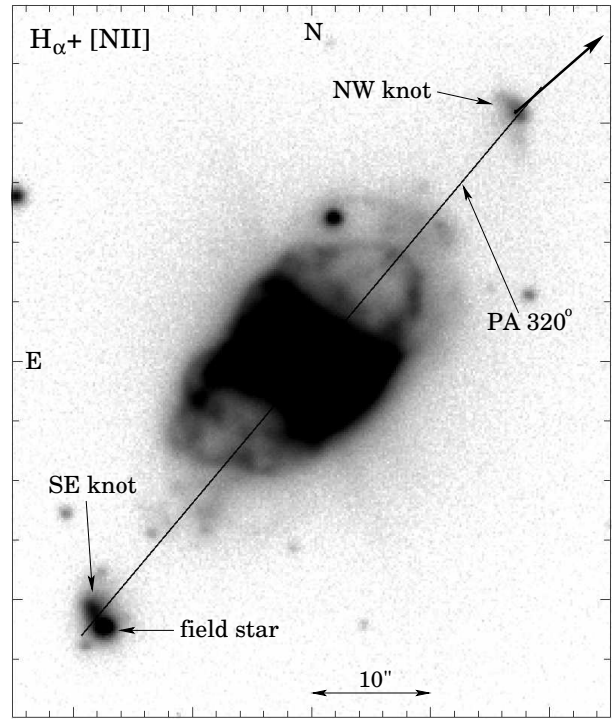}
   \caption{Grey-scale reproduction of the 2008.67 H$\alpha$+[N\,{\sc ii}] image of
     Hu\,1-2. The grey levels are logarithmic. The NW and
     SE knots are indicated as well as a field star partially superimposed on
     the SE knot. The slit
     position used for high resolution spectroscopy is drawn and labeled by
     its position angle (slit width not to scale). The arrow at the NW knot indicates the shift (10$''$) and its direction 
     (PA 312$^{\circ}$) expected in 500\,yr, according to the proper motion
     deduced for this knot (see text). North is up, east to the left, and 
     the scale is indicated.}
    \end{center}
\end{figure}

\begin{itemize}

\item Epoch\,1: 1951.51. Image from the POSSI-E, plate 08HM, obtained on
  1951 June 5 with the Palomar Schmidt telescope on a 103aE plate with a plexi filter. The exposure time was 50 minutes. 
  It is digitized with a plate scale of 1$\farcs$70\,pixel$^{-1}$ and
  its spatial resolution (FWHM of field stars) is $\simeq$ 3$\farcs$5.

\item Epoch\,2: 1953.68. Image from the POSSI-E, plate 089E, obtained on 1953
  September 5 with the Palomar Schmidt telescope, on a 103aE plate with a plexi filter. The exposure time was 45 minutes. 
  The plate scale is 1$\farcs$70\,pixel$^{-1}$ and its spatial resolution is $\simeq$ 3$\farcs$5.

\item Epoch\,3: 1987.56. Image from the POSSII-F, plate A1I4, obtained on 1987
  July 26 with the Oschin Smidth Telescope on a IIIaF plate with a RG610 filter. The exposure time was 75 minutes. 
  The plate scale is 1$\farcs$01\,pixel$^{-1}$ and its spatial resolution is $\simeq$ 3$\farcs$0.

\item Epoch\,4: 1994.54. Image from the IAC Morphological Catalog of Northern Galactic PNe (Manchado et al. 1996)
  obtained on 1994 July 24 with the IAC80 telescope (Observatorio del Teide, Tenerife), on a 1k$\times$1k CCD Thomson with 
  an H$\alpha$+[N\,{\sc ii}] filter (FWHM $\simeq$ 50{\AA }). The exposure time was 3600 seconds. 
  The plate scale is 0$\farcs$435\,pixel$^{-1}$ and its spatial resolution is $\simeq$ 2$\farcs$2.

\item Epoch\,5: 2008.67. H$\alpha$ and [N\,{\sc ii}] images obtained on 2004
  July with ALFOSC\footnote{ALFOSC (Andalucia Faint Object Spectrograph and Camera) is provided by the Instituto de Astrof\'{\i}sica de 
Andaluc\'{\i}a (IAA) under a joint agreement with the University of Copenhagen and NOTSA.} at the Nordic Optical Telescope 
(NOT, Roque de los Muchachos, La Palma), on a 2k$\times$2k  EEV CCD. We use the IAC narrow-band H$\alpha$ (FWHM $\simeq$ 8\,{\AA }) and 
  [N\,{\sc ii}] (FWHM $\simeq$ 9\,{\AA }) filters. The exposure time was 900\,s for each filter. The plate scale is 
  0$\farcs$19\,pixel$^{-1}$ and its spatial resolution is 0$\farcs$85. The individual H$\alpha$ and [N\,{\sc
    ii}] frames were co-added to produce the H$\alpha$+[N\,{\sc ii}] image of
  epoch\,5 shown in Figure\,1.

\end{itemize}

The images were registered with routines within the MIDAS package, using the
2008.67 image as reference, by means of 11 faint field stars that do not show noticeable 
proper motions in the $\simeq$ 57 yr time baseline. After the registering process, the
positions of the 11 fiducial stars (defined by their centroid) in the 2008.67 epoch are within $<$ 0$\farcs$1 their 
positions in the 1951.51 and 1953.68 epochs and within $<$ 0$\farcs$06 their positions in the 1987.56 and
1994.54 epochs. The resulting intrinsical error in the proper motion
between the different epochs ($<$ $\pm$ 0.9\,mas\,yr$^{-1}$) is negligible and
does not affect the measurement of nebular proper motions.

\begin{figure}
   \begin{center}
   \includegraphics[width=45mm,clip=]{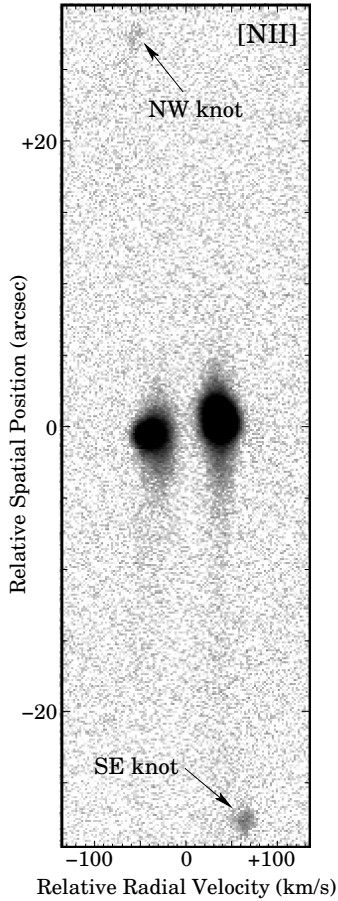}
  \caption{Grey-scale position-velocity map of the [N\,{\sc ii}]$\lambda$6583
    emission line along PA 320$^{\circ}$ (see Fig.\,1). Grey levels are 
logarithmic. Emission features from the NW and SE knots are indicated. The origin (0,0) corresponds to the spatial 
and radial velocity centroid of the relatively bright emission features from the inner nebular regions.} 
 \end{center}
\end{figure}

\subsection{High resolution spectroscopy}

A high resolution long-slit spectrum of Hu\,1-2 was obtained with
IACUB\footnote{The IACUB uncrossed echelle spectrograph was built in a
  collaboration between the IAC and the Queen's University of Belfast} at the
Nordic Optical Telescope (NOT) on Roque de los Muchachos Observatory (La
Palma) on 2004 June. We used a 1k$\times$1k Thompson CCD and a narrow--band filter to isolate the
H$\alpha$ and [N\,{\sc ii}]$\lambda$$\lambda$6548,6583 
emission lines. The slit (0$\farcs$65 wide) was oriented at position
angle (PA) 320$^{\circ}$, the orientation of the knots of Hu\,1-2 (see
below). The orientation of the slit is also shown in Fig.\,1. Exposure time was 900\,s. The
spectral resolution (FWHM) is 8 km\,s$^{-1}$ and the seeing was $\simeq$
1$''$. The spectrum was reduced using standard procedures for long-slit
spectroscopy within the IRAF package. Figure\,2 presents a position-velocity
map of the [N\,{\sc ii}]$\lambda$6583 emission line obtained from this spectrum. 

\section{Results and discussion}

The image in Fig.\,1 shows Hu\,1-2 consisting of an elliptical/bipolar main shell of 
$\simeq$ 14$''$$\times$24$''$ in size and oriented at PA $\simeq$ 320$^{\circ}$, and reveals 
several outer structures around the polar regions of the main shell (see Miranda et al. 2011 for 
a description of the morphological components in Hu\,1-2). The northwestern knot hinted by \citet{b4} is 
clearly detected in the 2008.67 image. The southeastern counterpart is identified in
this image for the first time owing its subarcsecond spatial resolution that permits to
separate the knot from a very close field star. We will refer to them as the
NW and SE knots. They present bow-shock-like morphology
with a central emission peak and extended wings, are located at 27$\farcs$5 from the
central star of Hu\,1-2 (Heap et al. 1990: Miranda et
al. 2011), and oriented at PA 320$^{\circ}$ that coincides with the
orientation of the main nebular axis. The SE knot is brighter than the NW one. 

\begin{figure}
    \begin{center}
    \includegraphics[width=80mm,clip=]{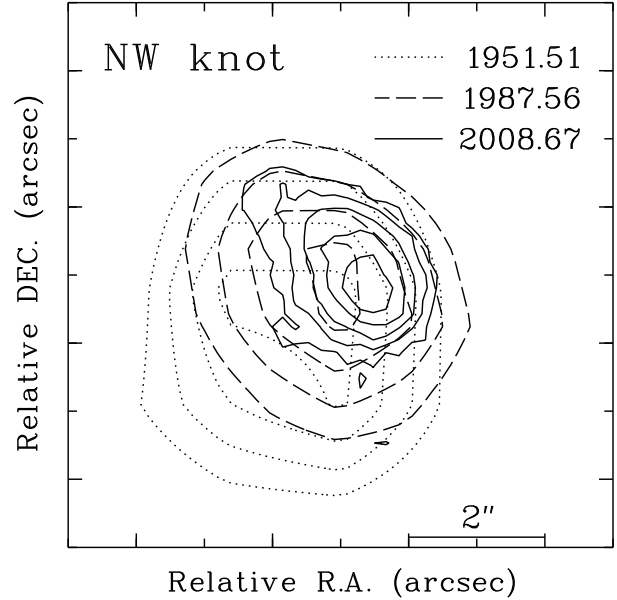}
  \caption{Intensity contour plots of the NW knot derived from the images of
    the 1951.51, 1987.56 and 2008.67 epochs after registering. The contours
    are arbitrary and have been chosen to highlight the regions of the knot
    around the intensity peak. North is up, east to the left. } 
 \end{center}
\end{figure}

Emission features from the bipolar knots are detected in the long-slit
spectrum (Fig.\,2) with radial velocities of $\pm$ 60\,km\,s$^{-1}$ (NW knot
blueshifted, SE knot redshifted). The inclination angle of the knots is
unknown. From an analysis of high-resolution, long-slit H$\alpha$ spectra 
at several PAs, Miranda et al. (2011) obtained an upper limit 
of 10$^{\circ}$ with respect to the plane of the sky for the inclination of the 
main nebular axis of Hu\,1-2 (see also Miranda et al. 2012 in preparation). If the knots moved along
the main nebular axis, as suggested by the coincidence of the orientations,
their expansion velocity would be $>$ 340\,km\,s$^{-1}$. It should be noted that a
more accurate value for the expansion velocity cannot be
obtained mainly because of the very small inclination angle of the object (see below). In any case, 
these results conclusively demonstrate that the NW and SE knots 
constitute a high velocity, collimated bipolar outflow. Moreover, the morphological and kinematical properties 
point out that the NW and SE knots represent the working surfaces of high-velocity bullets. 
We also note that the expansion velocity of these knots 
is high for collimated outflows in PNe (see Guerrero et al. 2002) and only in a few cases expansion velocities 
$>$ 300\,km\,s$^{-1}$ are observed (e.g., Riera et al. 2003; Guerrero \& Miranda 2011)

Figure\,3 shows intensity contour plots of the NW knot derived from the images
of the 1951.51, 1987.56 and 2008.67 epochs after registering. In spite of the different spatial
resolution in the images, a preliminary inspection of Fig.\,3 shows that the intensity
peak of NW knot has shifted towards the northwest in the last
$\simeq$ 57 years. This is corroborated by the 1953.68 and 1994.54 images (not
shown in Fig.\,3 for clarity). A quantitative
measurement of the position of the NW knot (defined by its centroid) was obtained
by fitting a two dimensional Gaussian line profile to the intensity distribution
of the knot in each epoch. Figure\,4 presents the relative $\alpha$ and $\delta$ positions of the NW knot 
as a function of time (epoch). Least-squares fits to the data in Fig.\,4 yield
a proper motion ($\mu$) $\mu$$_{\alpha}$ = $-$15.5 $\pm$ 5.0 mas\,yr$^{-1}$ 
and $\mu$$_{\delta}$ = $+$13.1 $\pm$ 3.1 mas\,yr$^{-1}$. By combining these
values and their errors, we obtain a proper motion for the NW knot $\mu$ =
20 $\pm$ 6 mas\,yr$^{-1}$ along PA = 312$^{\circ}$ $\pm$ 15$^{\circ}$. 
The direction of the proper motion vector agrees quite well with the
orientation of the main nebular axis and bipolar knots, strengthening the idea
that the collimated outflow move along that axis. Assuming ballistic ejection and origin 
at the central star, we obtain a dynamical age of 1375$^{+590}$\llap{$_{-320}$}\,yr for the NW knot.

\begin{figure}
    \begin{center}
    \includegraphics[width=80mm,clip=]{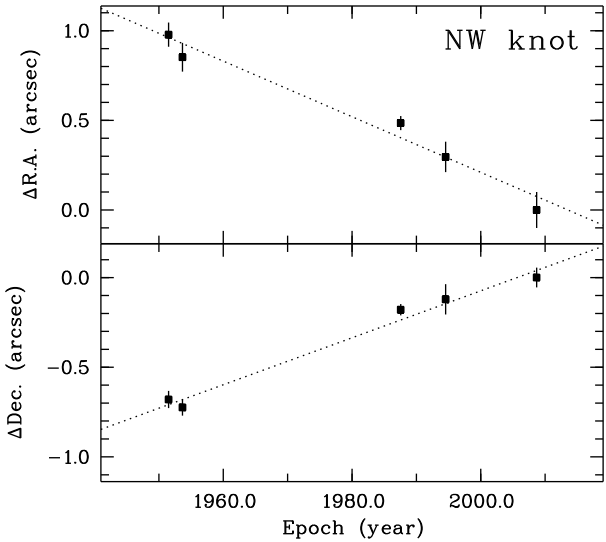}
  \caption{Position of the centroid of the NW knot versus time, relative
    to the position in the 2008.67 epoch. Error bars are 3$\sigma$ values. The
    dotted lines represent least-squares fits to the data.} 
 \end{center}
\end{figure}

A similar proper motion analysis cannot be carried out for the SE knot as it has been only detected 
in the 2008.67 image. However, the spatio-kinematical
properties (morphology, distance to the central star and radial velocity) of the SE
knot are virtually identical to these of the NW knot. Therefore, it can be 
inferred that other properties (proper motion vector, origin, age) are also identical in
both knots, too.

From the expansion velocity of 340\,km\,s$^{-1}$ and the proper motion of 20\,$\pm$\,6\,mas\,yr$^{-1}$ of the knots, we obtain a 
distance of 3.5$^{+1.5}$\llap{$_{-0.8}$}\,kpc to Hu\,1-2, with a lower limit of 2.7\,kpc giving the 
uncertainty in $\mu$ and that 340\,km\,s$^{-1}$ is a lower limit to the expansion velocity. As already mentioned, the very 
small inclination angle (from the sky) of the NW and SE knots represents a problem to obtain a more 
accurate value for the expansion velocity and, hence, for the distance. In Hu\,1-2, the assumption that the
NW and SE knots move along the main nebular axis is strongly supported by the
coincidence of their orientations and the direction of the proper motion. As for the inclination of the 
main nebular axis, Miranda et al. (2011) obtain an upper limit of 10$^{\circ}$ from the sky, being  
inclination angles of, say, 2$^{\circ}$, 5$^{\circ}$ or 7$^{\circ}$ also compatible with the observed 
spatio-kinematical properties of the main shell. An uncertainty of a few degrees in the inclination angle is extremely critical 
in the determination of the expansion velocity for inclination angles $<$ 10$^{\circ}$. For instance, 
if the NW and SE knots move at 5$^{\circ}$ from the sky (only 5$^{\circ}$ less than
the upper limit of 10$^{\circ}$), their 
expansion velocity is 688\,km\,s$^{-1}$ and the lower limit for the distance
results to be 5.6\,kpc, a factor $>$ 2 larger than the value deduced for
an angle of 10$^{\circ}$. Unless the expansion velocity (or inclination angle) 
of the NW and SE knots can be very much constrained by some method, the proper motion method 
will provide a lower limit for the distance only. 

It is interesting to compare the lower limit of 2.7\,kpc with other distance estimates to
Hu\,1-2. Statistical distances range between 1.3\,kpc (Amnuel et al. 1984) and
5.6\,kpc (Cahn 1976; see also Acker et al. 1992, Phillips 2004, and
  Stanghellini \& Haywood 2010). Pottasch (1983) and Sabbadin (1986) 
obtained an individual distance of 1.5--1.6\,kpc by the extinction method. Hajian \& Terzian (1996) set a 
lower limit of 1.2 kpc from the non detection of angular expansion in VLA 6\,cm radio continuum data at two epochs. 
The lower limit of 2.7\,kpc rules out a large fraction of the statistical
distances. For statistical distances $>$ 2.7\,kpc (e.g., Cahn 1976; Stanghellini \&
Haywood 2010) the uncertainty in the
expansion velocity does not allow us to drawn firm conclusions from possible
coincidences. In general, this result points out that 
statistical distances based on assumptions about nebular properties should be
used with caution when analyzing individual PNe (see Frew 2008 and Tafoya et
al. 2011). Furthermore, the extinction distance is 
also excluded, most probably because Hu\,1-2, at a galactic latitude of
$\simeq$ $-$8$\fdg$8, is located outside 
the interstellar reddenig layer and the extinction method provides a too small distance (see 
Phillips 2006). 

As shown above, proper motions of collimated outflows in PNe, if detected in
POSS red plates, can be obtained by comparing these plates with recent
H$\alpha$+[N\,{\sc ii}] CCD images. This is due to the very peculiar optical spectrum  
of collimated outflows in PNe, which is entirely dominated by H$\alpha$ and [N\,{\sc ii}] emission lines and excludes 
a noticeable contribution from other lines in the emission detected in the POSS red plates. As for now, collimated 
outflows in POSS red plates have been identify in NGC\,7009, KjPN\,8 and Hu\,1-2 (see above) but a careful inspection of 
these plates is desirable to identify more cases and to attemp a measurement of their proper motion. These, in 
conjunction with spatio-kinematical models of the collimated outflows and/or nebular shell, may provide a powerful method 
to obtain (or to constrain) distances to PNe. This procedure is not restricted to collimated outflows 
only but can also be applied 
to other high velocity features (knots, filaments) in PNe if their emission in the red spectral range is
dominated by the H$\alpha$ and [N\,{\sc ii}] lines. In this respect it is worth noting that
Meaburn et al. (2008) used an 1956 SAAO red plate and a 2007 H$\alpha$+[N\,{\sc
  ii}] CCD image to successfully measure the proper motion of a large number of
knots in the western lobe of NGC\,6302 and to obtain its distance.  While a shortcoming of the
POSS red plates is their low spatial resolution (e.g., the case of the SE
knot), clear advantages of these plates are the large time baseline ($>$ 50\,yr) 
and the possibility of using more than two epochs (including POSSI and POSSII) 
to determine the proper motions.

\section{Conclusions}

We have presented a subarsecond H$\alpha$+[N\,{\sc ii}] image and a 
high resolution long-slit spectrum of Hu\,1-2 covering its bipolar knots. In
addition, we have analyzed five images of Hu\,1-2 with a time baseline of
$\simeq$ 57 yr, including three POSS red plates in which one
of the bipolar knots can be identified, to measure its proper motion and to
constrain the distance to the nebula. The main conclusions of this paper are:

\begin{itemize}

\item The subarcsecond image allows us to identify the two bipolar knots of
  the pair, while in previous images only one of them is clearly detected. The
  knots present bow-shock-like morphology, are located at 27$\farcs$5 from
  the central star and oriented at PA 320$^{\circ}$ that coincides
  with the orientation of the main nebular axis.

\item The radial velocity of the knots is $\pm$ 60 km\,s$^{-1}$. If the knots moved
  along the main nebular axis (tilted $<$ 10$^{\circ}$ with respect to the
  plane of the sky), their expansion velocity would be $>$ 340\,km\,s$^{-1}$. Therefore,
  morphology and kinematics demonstrate that these knots constitute a true high
  velocity, collimated bipolar outflow and, most probably, represent bow-shocks associated 
  to high velocity bullets. 

\item A proper motion of 20 $\pm$ 6 mas\,yr$^{-1}$ along PA 312$^{\circ}$
  $\pm$ 15$^{\circ}$ is obtained for the bipolar knots. The corresponding
  dynamical age is 1375$^{+590}$\llap{$_{-320}$}\,yr. 

\item A lower limit of 2.7\,kpc for the distance to Hu\,1-2 is required to
  make compatible the measured proper motion with the spatio-kinematical
  properties of the bipolar outflow. This lower limit rules out a large fraction of the statistical 
  distances and the extinction distance previously determined for Hu\,1-2. 

\end{itemize}

\section*{Acknowledgments}

We thank our anonymous referee for comments that have improved the presentation and 
discussion of the results. We are grateful to the group of support astronomers of the IAC for making us
available the narrow-band filters used in the NOT observations. LFM, MB, and MAG acknowledge 
support by grant AYA2008-01934 of the Spanish MCINN (co-funded by FEDER funds). AR acknowledges 
support by grant AYA2008-06189-C03 of the Spanish MICINN (co-funded with FEDER funds). LFM acknowledges 
support from grant IN8458-2010/061 of Xunta de Galicia (co-funded by FEDER funds). The
Digitized Sky Surveys were produced at the Space
Telescope Science Institute under U.S. Government grant NAG\,W-2166. The
images of these surveys are based on photographic data obtained using the
Oschin Schmidt Telescope on Palomar Mountain and the UK Schmidt Telescope. The
plates were processed into the present compressed digital form with the
permission of these institutions. The National Geographic Society - Palomar
Observatory Sky Atlas (POSS-I) was made by the California Institute of
Technology with grants from the National Geographic Society. The second
Palomar Observatory Sky Atlas (POSS-II) was made by the California Institute
of Technology with funds from the National Science Foundation, the National
Geographic Society, the Sloan Foundation, the Samuel Oschin Foundation, and
the Eastman Kodak Corporation. The Oschin Schmidt Telescope is operated by the
California Institute of Technology and Palomar Observatory. The UK Schmidt
Telescope was operated by the Royal Observatory Edinburgh, with funding from
the UK Science and Engineering Research Council (later the UK Particle Physics
and Astronomy Research Council), until 1988 June, and thereafter by the
Anglo-Australian Observatory. Supplemental funding for sky-survey work at the
ST\,ScI is provided by the European Southern Observatory.

\end{document}